\documentclass[manuscript,12pt]{aastex}

\newcommand{\com}[1]{{}}
\newcommand{\rev}[1]{{#1}}
\newcommand{\revs}[1]{{#1}}

\title{Small Planetesimals in a Massive Disk Formed Mars 
}

\author{Hiroshi Kobayashi}
\affil{
Department of Physics, Graduate School of Science, 
Nagoya University, Furo-cho, Chikusa-ku, Nagoya 464-8602, Japan 
}
\email{hkobayas@nagoya-u.jp}
\affil{
}

\affil{}
\affil{}

\author{Nicolas Dauphas}

\affil{
Origins Laboratory, Department of the Geophysical Sciences and Enrico Fermi Institute, The University of Chicago, 5734 South Ellis Avenue, Chicago, Illinois 60637, USA
}
\email{dauphas@uchicago.edu}

\affil{}
\affil{}
\affil{}
\affil{}
\affil{{\bf Icarus, 225, 122}}
\affil{}

\affil{
}
\affil{}
\affil{}

\begin{document}

\begin{abstract}

Mars is likely to be a planetary embryo formed through collisions with
 planetesimals, which can explain its small mass and rapid formation
 timescale obtained from $^{182}$Hf-$^{182}$W chronometry.  In the
 classical theory of planet formation, the final embryo mass is
 determined only by the solid surface density.  However, embryos can
 stir surrounding planetesimals, leading to fragmentation through
 erosive (cratering) collisions.  We find that radial drift of small
 fragments can drastically reduce the solid surface density.  On the
 other hand, embryo growth is accelerated by fragment accretion. Since
 collisional fragmentation efficiency depends on the initial size of
 planetesimals, the final embryo mass and its growth time are determined
 by the initial planetesimal size and disk surface density.  We have
 investigated the effect of these two parameters on the mass of Mars and
 the predicted radiogenic excess of $^{182}$W in the martian mantle. Two
 scenarios can explain the rapid formation of small Mars: (i) it formed
 by accretion of small planetesimals in a massive disk or (ii) it formed
 from large planetesimals but its growth was arrested by the inward then
 outward migration of Jupiter. Taking into account all constraints, we
 conclude that Mars is likely to have formed in a massive disk of about
 $\sim 0.1$ solar mass from planetesimals smaller than $\sim 10$\,km in radius. Such
 small planetesimal size cannot explain core accretion of Jupiter,
 suggesting that there may have been a heliocentric gradient in
 planetesimal size in the solar nebula.

\end{abstract}



\section{INTRODUCTION}

Mars, one of four terrestrial planets in the solar system, has a small mass of 
$0.11$ $M_\oplus$ ($M_\oplus=$Earth mass)
and an orbital semimajor
axis of 1.5\,AU. 
Terrestrial planets are believed to have formed via collisions among
Mars-sized planetary embryos in the long-term orbital instability that
followed substantial gas dissipation 
\citep[{\it i.e.}, chaotic growth; ][]{chambers96,iwasaki,kominami,kenyon06,ogihara}. 
The $^{182}$Hf-$^{182}$W short-lived chronometer ($t_{1/2}=$ 9\,Myr)
gives a very short accretion timescale for Mars \citep{dauphas}, similar
to the gas dissipation time inferred from disk and meteorite observations 
\citep{dauphas2,haisch}. This indicates that Mars is likely to
be a remnant of the early generation of planetary embryos formed via
collisional accretion of planetesimals. 

Collisions between planetesimals produced planetary embryos, which grew
further through collisions with surrounding planetesimals
\citep{wetherill93,weidenschilling97,kokubo98}.  Once embryos became
massive, they stirred surrounding planetesimals, producing collisional
fragments that contributed to embryo growth
\citep{wetherill93,inaba03,chambers06,chambers08,kenyon04,kenyon09,ormel_kobayashi}.
When the surrounding bodies were depleted due to the radial drift of
small fragments and/or accretion onto embryos, the growth of embryos
stalled \citep{kobayashi10}.  In oligarchic growth, the initial size of
planetesimals and disk mass influence the final masses of embryos and
their accretion timescales. Small planetesimals are easily fragmented
during collisions. The resulting fragments are small enough to feel some
drag from the gas, leading them to spiral towards the Sun. As a result,
small planetesimals tend to produce embryos that have smaller masses
than when large initial planetesimals are considered. On the other
hand, the interaction of fragments with the gas damps their relative
velocities and increases the efficiency with which they can be accreted
by embryos through gravitational focusing. Thus, with small
planetesimals the timescale of embryo growth is shorter than when large
planetesimals are considered.

During formation of the terrestrial planets, the segregation of
metal from silicate produced metallic cores, the timing of which can be
estimated using $^{182}$Hf-$^{182}$W systematics.  Hafnium, a lithophile
(i.e., rock-loving) element, was retained.  In contrast, tungsten, a siderophile (i.e.,
metal-loving) element, was partitioned into metal
cores. Hafnium-182 ($^{182}$Hf) decays to tungsten-182 ($^{182}$W) with a
half life of 9\,Myr. The decay of $^{182}$Hf in the martian mantle after
removal of W in the core resulted in the production of excess radiogenic
$^{182}$W compared to other isotopes of W in chondrites. The $^{182}$W
excess in the martian mantle is known from measurements of
shergottite-nakhlite-chassignite (SNC) meteorites that are thought to
have been ejected from Mars by impactors \citep[{\it
e.g.},][]{kleine04,kleine,foley,dauphas}. Strictly speaking, $^{182}$W
variations record the timing of core formation but this process
presumably tracked planetary accretion. In this contribution, we use
$^{182}$W excess in the martian mantle and the mass of Mars to constrain
the disk surface density and initial planetesimal size in the Mars-forming region.

\section{MARS FORMATION}
\label{141207_4Sep12}


We use the following disk model for the initial surface mass density of
planetesimals $\Sigma_{\rm s,0}$ and gas $\Sigma_{\rm g,0}$: 
\begin{eqnarray}
 \Sigma_{\rm s,0} &=& x
\Sigma_{\rm MMSN,s} \left(\frac{a}{1{\rm AU}}\right)^{-3/2}
  \, {\rm g\, cm}^{-2},\label{201002_7Dec12} \\
\Sigma_{\rm gas,0} &=& x
\Sigma_{\rm MMSN,g} \left(\frac{a}{1{\rm AU}}\right)^{-3/2} 		     
  \, {\rm g\, cm}^{-2},\label{eq:sigma_gas} 
\end{eqnarray}
where $\Sigma_{\rm MMSN,s} = 7.1 \,{\rm g\,cm^{-2}}$ and $\Sigma_{\rm
MMSN,g} = 1.7 \times 10^3 \,{\rm g\,cm^{-2}}$ are, respectively, 
the solid and gas surface densities at 1\,AU in the minimum-mass solar 
nebula (MMSN) model \citep{hayashi}, and $a$ is the distance from the Sun.  We vary the scaling factor $x$ to
investigate the conditions for Mars formation.  The solid surface density
evolves naturally in the simulation, while the gas density is artificially imposed to decrease as a function of time as $\exp(-t/\tau_{\rm d})$ with a gas
dissipation timescale $\tau_{\rm d}$. 
Disk observations indicate that $\tau_{\rm d}$ is on the order of
several million years 
\citep{haisch}. 

\citet{kobayashi10,kobayashi11} showed that model outputs of statistical
 simulations of runaway and oligarchic growth are controlled primarily
 by the planetesimals that dominate the initial solid surface density,
 so the results are largely insensitive to the initial size distribution
 of other minor bodies. For this reason, we have adopted a fixed initial
 planetesimal radius $r_0$ rather than using an actual initial size
 distribution, which is poorly known anyhow
 \citep{morbidelli,weidenschilling11}.
We limit our analysis to $r_0 \geq 1$\,km because in a turbulent disk,
 smaller planetesimals would not be sensitive to gravitational focusing, thus preventing runaway growth to proceed \citep[e.g.,][]{youdin,ormel_cuzzi, ormel}. 

We performed numerical simulations of Mars formation via collisional
evolution of planetesimals. In the model, the disk is divided into four
annuli. The width of each annulus is given by 0.2 times the annulus'
characteristic radius defined by the mean of the inner and outer radii
of the annulus.  The characteristic radii of the annuli are set to 1.5,
1.8, 2.2, and 2.7\,AU.  \citet{kobayashi10,kobayashi11} showed that
because the timescale for dust removal by gas drag is much shorter than
the timescale for embryo growth, considering a larger disk did not
change significantly the final embryo mass.  The
mass distribution of bodies in each annulus is treated as discrete mass
batches with the mass ratio between adjacent mass batches fixed at 1.2.  We
follow the evolution of the mass distribution via collisions between
bodies. The collisional rates depend on the relative velocities between
colliders, which depend on their eccentricities and inclinations.  The
mass distribution affects eccentricities and inclinations by
collisionless interaction between bodies.  We therefore calculate the
mass and velocity evolution simultaneously, applying the collision rates
\citep{inaba01} and the velocity evolution rates due to mutual
interactions of bodies \citep{ohtsuki02} and due to gas drag
\citep{adachi76}.  We do not treat collisions and collisionless
interactions between bodies in different annuli and therefore adopt an
annulus width larger than the feeding zones of planetary embryos.
The total fragment mass ejected by a single impact between colliders
with masses $m_1$ and $m_2$ is assumed to be $(m_1 + m_2) \phi/(1+\phi)$
using the dimensionless impact energy $\phi = m_1 m_2 v^2 /2 (m_1+m_2)^2
Q_{\rm D}^*$ \citep{KT10, kobayashi10}, where $v$ is the collisional
velocity between the colliders and $Q_{\rm D}^*$ is the specific impact
energy (${\rm erg \, g}^{-1}$ of impacted material) needed for the ejection of a
half-mass of colliders.  Hydrodynamic impact simulations give $Q_{\rm
D}^*$ for 100 km sized or smaller bodies \citep{benz}, while $Q_{\rm
D}^*$ of bodies with radius $r \ga 100$\,km are determined by the
gravitational binding energy \citep{stewart}. Connecting the two regimes
(above and below about 100\,km), we express $Q_{\rm D}^*$ value of a
body with mass $m$, radius $r$, and density $\rho$ as $Q_{\rm D}^* =
Q_{\rm 0s}(r/1\,{\rm cm})^{\beta_{\rm s}}+Q_{\rm 0g} \rho (r/1\,{\rm
cm})^{\beta_{\rm g}} + 2 C_{\rm gg} G m /r$, where $G$ is the
gravitational constant and $C_{\rm gg}=9$ from collisional simulations
with distinct element method \citep{stewart}. Here we use $Q_{\rm
0s}=3.5\times 10^7{\rm erg \, g}^{-1}$, $Q_{\rm 0g} = 0.3\, {\rm erg\, cm}^3 \,
{\rm g}^{-2}$, $\beta_{\rm s} = -0.38$, and $\beta_{\rm g} = 1.36$,
which were obtained from hydrodynamic simulations for basalt with $v=3
\,{\rm km/s}$ \citep{benz}. \com{$Q_{\rm D}^*$ depends on the materials
and structures of bodies and collisional velocities, however the
dependence has still uncertainty.  The $Q_{\rm D}^*$ value we used is
larger than rabble piles. Our planetesimals produce relatively large
embryos. However, $Q_{\rm D}~*$ is determined by pure gravity for $r_0
\ga 100$\,km. The uncertainty is minor to rule out $r_0 \ga 100$\,km.}
Following \citet{wetherill93} and \citet{inaba01}, when the bodies reach
a certain mass $m_{\rm run}$ such that the sum of their mutual Hill
radii equals the radial-bin width divided by $\tilde b = 10$
\citep{kokubo}, the bodies are regarded as ``runaway bodies'' that do
not experience collisions and dynamical interactions due to close
encounters among them. Using statistical methods, we can accurately
calculate embryo formation and growth until the subsequent stage of chaotic growth, which
may not be significant for Mars. The statistical simulation used in the
present contribution reproduces the results
of $N$-body simulations when collisional fragmentation is ignored
\citep{kobayashi10}.  It also reproduces the analytical solution for
mass loss due to a collisional fragmentation cascade \citep{KT10}.

Embryos grow via collisions with surrounding planetesimals and
fragments. 
If we assume that the bodies surrounding embryos are not lost at
all, 
then embryos can accrete all the bodies in their feeding zones. The maximum mass of the
embryo is called the isolation mass and is given by \citep[e.g.,][]{kokubo} 
\begin{equation}
 M_{\rm iso} = 0.13 \, x^{3/2}
\left(\frac{a}{1.5\,{\rm AU}}\right)^{3/4}
M_\oplus,  
\end{equation}
where we use Eq.~(\ref{201002_7Dec12}) for the solid surface density.
Therefore, a disk with $x\sim 1$ produces Mars-size embryos at 1.5\,AU if embryos
can collect all solids. \com{The definition of the MMSN is slightly
different: The isolation mass does not need to correspond to the masses
of the terrestrial planets. For example, Earth and Venus cannot reach
their masses prior to the chaotic growth.}  However, since small
planetesimals are vulnerable due to low gravity, mass loss can occur by
collisional fragmentation of planetesimals with radial
drift of the resulting fragments, so the final embryo mass can be smaller
than $M_{\rm iso}$.  Destructive collisions between planetesimals
produce small bodies, which can in turn be broken apart resulting in
still smaller bodies.  In the collisional fragmentation cascade,
the mass distribution of bodies follows a power law; the surface number
density of bodies with mass ranging from $m$ to $m+dm$ is proportional
to $m^{-\alpha} dm$, where $\alpha = (11+3p)/(6+3p)$ with $v^2/Q_{\rm
D}^* \propto m^{p}$ \citep{KT10}.  
The mass distribution is determined by the collisional cascade for
bodies with $r \la r_0$, while runaway growth of planetesimals
regulates the mass distribution of bodies with $r \ga r_0$. 
With such a mass distribution,
planetesimals with radii $\sim r_0$ determine the solid surface density.  The
timescale of planetesimal depletion by the collisional fragmentation
depends on $Q_{\rm D}^*$ of planetesimals and is given by
\citep{KT10} 
\begin{eqnarray}
 \tau_{\rm cc} &=&1.1 \times 10^4 \,\, x^{-0.71}
  \left(\frac{r_0}{10\,{\rm km}}\right)^{1.69}
  \left(\frac{a}{1.5\,{\rm AU}}\right)^{3.2}
  \nonumber
  \\ && \times
  \left(\frac{M}{0.11\,M_\oplus}\right)^{-0.48} 
  \left(\frac{Q_{\rm 0g}}{0.3 \, {\rm erg\, cm}^3{\rm
   \,g}^{-2}}\right)^{-0.72} {\rm yr},  
\end{eqnarray}
where we assume that the planetesimal eccentricities are in an equilibrium
state controlled by stirring by embryos and damping by gas drag and use $Q_{\rm D}^* = Q_{\rm 0g} \rho (r_0/1\,{\rm cm})^{\beta_{\rm
g}}$ with $\rho = 3 {\rm g\, cm}^{-3}$ for planetesimals. 
Erosive (cratering) collisions with specific impact energies smaller
than $Q_{\rm D}^*$ control
this timescale \citep{KT10}, which is 4--5 times shorter than that derived by ignoring erosive
collisions as previous studies did \citep[e.g.,][]{wetherill93}. This timescale was obtained by assuming power law mass distribution in the collision cascade. Several effects can lead to departure from such power-law behavior but this would not change significantly the timescale because it is determined
by bodies for which the mass distribution does not deviate considerably from
a power-law distribution with $\alpha \sim 11/6$.
Since $\tau_{\rm cc}$ is inversely proportional to the solid surface
density \citep{KT10}, $\tau_{\rm cc}$ increases with decreasing
$\Sigma_{\rm s}$ and the full-depletion timescale of planetesimals 
is much longer than
$\tau_{\rm cc}$. 

The timescale for embryo growth via planetesimal accretion is given by \citep{kokubo,chambers06,kobayashi10}, 
\begin{eqnarray}
 \tau_{\rm grow,p} &=& 2.7 \times 10^5 \,\, x^{-7/5} 
  \left(\frac{r_0}{10\,{\rm km}}\right)^{2/5}
  \nonumber
  \\
 &&\times
  \left(\frac{a}{1.5\,{\rm AU}}\right)^{27/10}
  \left(\frac{M}{0.11\,M_\oplus}\right)^{1/3}
  {\rm yr}.  
\end{eqnarray}
For $r_0 \ga 100\,$km, embryos can reach $M_{\rm iso}$ in a disk with
$x\approx 1$ because $\tau_{\rm grow,p} \la \tau_{\rm cc}$. However, the
final embryo mass is smaller than the isolation mass for $r_0 \la
100$\,km.  Larger $x$ is needed for smaller planetesimals to 
compensate for loss of mass through gas drift of collision fragments and 
achieve the
same final embryo mass.  Note that the necessary $x$ to produce
Mars-size embryos is lower than $x$ given by posing $\tau_{\rm cc} = \tau_{\rm
grow,p}$, because $\tau_{\rm cc}$ and $\tau_{\rm
grow,p}$ increase when the solid surface density decreases 
\citep{kobayashi10}.

For small values of $r_0$ ($\la
1$--10\,km), embryo growth proceeds to a large extent by fragment
accretion.  The collisional fragmentation cascade halts when
eccentricities and inclinations of the bodies are damped well by gas
drag.  Fragments thus accumulate at $v^2 \approx 0.5 Q_{\rm D}^*$. 
This condition 
gives a typical fragment size of 1--10\,m depending on the embryo mass $M$.  The
radial drift timescale of such bodies is given by \citep[see the
derivation in][]{kobayashi10}
\begin{eqnarray}
 \tau_{\rm drift} &=& \frac{\tau_{\rm stop}}{2\eta}
= 
\frac{2^{1/3} \pi \tilde{b} Q_{\rm D}^* }{\eta v_{\rm k}^2 P_{\rm
VS,low} \Omega_{\rm K}}
\nonumber
\\
&=& 
6.9 \times 10^2 
  \left(\frac{a}{1.5 \, {\rm AU}}\right)
  \left(\frac{M}{0.11\,M_{\oplus}}\right)^{-1}
  \nonumber
  \\
 && \times 
  \left(\frac{Q_{\rm D}^*}{6.1\times 10^6 \,{\rm erg \, g}^{-1}}\right)
  {\rm yr}, 
\end{eqnarray}
where $\Omega_{\rm K}$ is the Keplarian frequency, we use the value of
$Q_{\rm D}^*$ for 
$r = 1$\,m, $\tau_{\rm stop}$ is the stopping time by
gas drag for the fragments under the Stokes regime, $v_{\rm K}$ is the
Keplarian velocity, $P_{\rm VS,low} = 73$ is the dimensionless viscous
stirring rate for low eccentricity, and $\eta = (v_{\rm K} - v_{\rm
gas})/v_{\rm K}$.  Since small fragments have low eccentricities and
inclinations, fragment accretion allows fast growth of embryos. 
The embryo growth timescale via fragment accretion is given by
\citep{kobayashi10}
\begin{eqnarray}
 \tau_{\rm grow,frag} &=& 4.3\times 10^3 \,\, x^{-1} 
  \left(\frac{M}{0.11\,M_\oplus}\right)^{1/3}
  \left(\frac{a}{1.5\,{\rm AU}}\right)^{2}
{\rm yr}.
\end{eqnarray}
If we ignore the radial drift of fragments, embryos reach $M_{\rm
iso}$ in a short timescale of $\tau_{\rm grow,frag}$ \citep{rafikov04,kenyon09}. 
However, fragments are
removed by radial drift due to gas drag in a shorter timescale of 
$\tau_{\rm drift}$, which stalls embryo growth before they reach
isolation masses. 
In the case of fragment accretion (small initial planetesimals), 
formation of Mars-size embryo requires $\tau_{\rm drift} \approx \tau_{\rm grow,frag}$; 
$x\ga 6$. 
With larger initial planetesimals, collisional fragmentation of
planetesimals is less effective and therefore embryos can reach the mass
of Mars for smaller $x$.

To summarize, small initial planetesimals are prone to fragmentation and 
mass loss occurs through gas-induced drift of fragments. As a result,
small planetesimals tend to produce small embryos (and vice versa). The
eccentricities and inclinations of fragments produced by collisions
between small planetesimals are damped by interaction with the gas. The
resulting reduction in relative velocities increases the gravitational
cross section of embryos for fragments, so small planetesimals allow
rapid embryo growth (and vice versa).

We start the investigation for embryo growth in a MMSN disk, where the
final embryo mass is about the mass of Mars if embryos can fully accrete
bodies in their feeding zones; $M_{\rm iso} \approx 0.1 M_{\oplus}$ at 1.5\,AU. Fig.~\ref{fig:embryo_growth}a shows the time evolution of
planetary embryo masses at 1.5\,AU for $r_0 = 1$, 3.7, 27 and $100\,$km
with $\tau_{\rm d} = 10$\,Myr. Growth rates and final masses of embryos
depend on $r_0$. Mars-sized embryos can be formed from planetesimals
with $r_0 \ga 10$\,km in the MMSN disk, while a more massive disk is
necessary to form Mars from smaller initial planetesimals.

The embryo mass evolution affects the predicted $^{182}$W radiogenic excess caused by the partitioning of W
into the core and the decay of $^{182}$Hf in the mantle. 
In the embryo mantle, 
the excess abundance of $^{182}$W compared to the
chondritic reference value (CHUR = chondritic uniform reservoir) is
given by 
\citep[see the derivation in Appendix \ref{appendix};][]{jacobsen,dauphas}
\begin{eqnarray}
 \varepsilon ^{182}{\rm W}_{\rm mantle} &=& q_{\rm W} 
  \left(\frac{^{182}{\rm Hf}}{^{180}{\rm Hf}}\right)
  f_{\rm mantle}^{\rm Hf/W} \, \lambda 
\nonumber\label{eq:eW}
\\
 && \times \int_0^{t} 
   \left(\frac{M (t^{\prime})}{M(t))}\right)^{1+f_{\rm 
   mantle}^{{\rm Hf/W}}} 
  \exp(-\lambda t^{\prime}) dt^{\prime}, 
\end{eqnarray}
where ${\varepsilon} ^{182}{\rm W}_{\rm mantle} = [(^{182}{\rm W} / ^{184}{\rm
W})_{\rm mantle} /(^{182}{\rm W} / ^{184}{\rm W})_{\rm CHUR} -1 ] \times 10^4$, $q_{\rm W} = (^{180}{\rm Hf}/^{182}{\rm
W})_{\rm CHUR} \times 10^4$, $f_{\rm mantle}^{{\rm Hf/W}} = ({\rm Hf}/{\rm W})_{\rm
mantle}/({\rm Hf}/{\rm W})_{\rm CHUR}-1$, 
$M(t)$ is the embryo mass at time $t$, 
and $\lambda$ is the decay constant of $^{182}{\rm Hf}$.  
From previous studies, 
$\varepsilon^{182}W_{\rm Mars\,mantle}
= 2.68 \pm 0.19$,  
$f_{\rm Mars\,mantle}^{{\rm Hf/W}} = 3.38
\pm 0.56$, $q_{\rm W} =1.07\times 10^4$ and
$(^{182}{\rm Hf}/^{180}{\rm Hf}) = (9.72 \pm 0.44) \times 10^{-5}$ \citep[][and references therein]{dauphas}. 
Our simulations generate as model output the time evolution of an embryo
mass $M(t)$, 
allowing us to calculate
$\varepsilon^{182}W_{\rm mantle}$. 
Fig.~\ref{fig:embryo_growth}b shows the evolution of
$\varepsilon^{182}{\rm W}_{\rm mantle}$ during
 embryo growth with different initial planetesimal radii in the MMSN disk.  
The predicted $\varepsilon^{182}{\rm W}_{\rm Mars\,mantle}$ values are smaller than
the measured one for $r_0 = 1$--100\,km.  
Small $r_0$ values produce $\varepsilon^{182}{\rm W_{\rm
mantle}}$ close to the martian mantle value
(Fig.~\ref{fig:embryo_growth}b) but the final embryo mass is too small
and a more massive disk is required (Fig.~\ref{fig:embryo_growth}a).

The results for massive disks with $x = 2.7$ are shown in Fig.~\ref{fig:embryo_growth_s2.7}. 
Smaller initial planetesimals with $r_0 = 3.7$\,km can make Mars-size
embryos, while larger planetesimals make 
embryos that are much larger than Mars. 
The predicted $\varepsilon^{182}$W value for $r_0 = 3.7$\,km is about
2.84, which matches the observation. 
Therefore, a parameter set of $x \approx 3$ and $r_0 \approx 4$\,km can
account for the mass and $\varepsilon^{182}$W value of Mars. 
For $x = 2.7$ and $r_0 = 3.7$\,km, the two inner annuli
produce 8 Mars mass bodies, while smaller bodies are formed in the outer
annuli close to the outer edge of the simulation. 
The predicted $\varepsilon^{182}$W values of the embryos are similar
but slightly lower for embryos formed in the outer disk relative to
those formed in the inner disk. 
In the proposed scenario, one of the 
embryos becomes Mars while the others accumulate to make Earth and Venus
size planets in the subsequent stage of chaotic growth (see \S \ref{sc:subsequent}).

To investigate the conditions required for Mars formation, we have
 evaluated systematically the effects of initial planetesimal size and
 solar disk mass on the final mass and accretion timescale of embryos in
 the Mars-forming region. Fig.~\ref{fig:cont_mass_ew}a summarizes the
 final embryo masses and the $\varepsilon^{182}W_{\rm mantle}$ values
 derived from 88 simulations at 1.5\,AU for 
a logarithmic grid of $x \times r_0$, where 
 $x = 1$, 1.4, 1.9, 2.7, 3.7, 5.2, 7.2, 10 and $r_0 = 1$, 1.9, 3.7,
 7.2, 14, 27, 52, 100, 190, 370, 720\,km. Final embryo masses reach the
 mass of Mars on the thick solid curve in the $x$ and $r_0$ space of
 Fig.~\ref{fig:cont_mass_ew}a.  Our statistical simulations present a
 mass distribution of embryos that spans a factor 4--5 in mass, which is
 a measure of uncertainty in the predicted embryo mass (thin solid
 curves).  The tendency that small initial planetesimals require a
 massive disk for the formation of Mars-sized bodies is explained by the
 analytical solutions for the final embryo mass derived by
 \citet{kobayashi10}.  Mass loss due to collisional
 fragmentation with radial drift of fragments reduces the final mass for
 small $r_0$.  The $\varepsilon^{182} {\rm W}_{\rm mantle} \pm \Delta
 \varepsilon^{182} {\rm W}_{\rm mantle}$ values are calculated from
 Eq.~(\ref{eq:eW}), where $\Delta \varepsilon^{182} {\rm W}_{\rm
 mantle}$ is the error in the predicted $\varepsilon^{182} {\rm W}_{\rm
 mantle}$ value calculated by propagating all uncertainties on model
 parameters.  At 1.5 AU, the parameter space that reproduces both the
 accretion timescale and final mass of Mars is $r_0 \la 10$\,km and $x
 \ga 1$.

To evaluate the robustness of our conclusion, we have 
carefully investigated  
the parameters that could affect the mass and accretion
timescale of Mars. In particular, the influences of embryo atmospheres
and gas drag on accretion dynamics, embryo migration, 
disk dissipation timescales, and disruption strength of bodies are discussed below.

Fig. \ref{fig:other_effects}a shows the results of simulations including
the collisional enhancement for capture of small particles by embryos 
induced by the presence of a planetary atmosphere, with a density profile determined by the
atmospheric opacity 
\citep{inaba_ikoma}.  The parameter space that explains the mass and
$\varepsilon ^{182}$W of Mars is similar to the case without the
collisional enhancement by planetary atmosphere.  The reason is that
embryos have small atmospheres and the atmospheric effect is
negligible for an embryo smaller than Mars \citep{kobayashi11}.

As discussed previously, gas can damp velocities of small particles,
thus increasing the efficiency of their capture by
embryos. \citet{ormel_klahr} further found that gas can directly affect the collisional
cross-sections between embryos and small bodies. While the dissipative
nature of gas drag promotes collision, strong particle-gas coupling acts
against it by allowing fragments to be carried along with the gas passed
the embryo.
Fig.~\ref{fig:other_effects}b
illustrates the results of simulations including the collisional cross
section affected by gas drag that is derived by \citet{ormel_klahr}. 
The collisional enhancement 
is significant only at $\tau_{\rm stop} \approx
\Omega_{\rm K}^{-1}$ for Mars-size embryos. Since the typical fragments have radii $\sim
10$\,m corresponding to $\tau_{\rm stop} \gg \Omega_{\rm K}^{-1}$ \citep{kobayashi10}, this collisional
enhancement is minor for Mars formation (compare 
Figs.~\ref{fig:cont_mass_ew}a with \ref{fig:other_effects}b). 

We additionally investigated how the gas dissipation 
time $\tau_{\rm d}$ affected the accretion history of Mars.  The final embryo mass and $\varepsilon ^{182}$W
value for $\tau_{\rm d} = 5$\,Myr are almost the same as
those for $\tau_{\rm d} = 10$\,Myr (Fig.~\ref{fig:other_td}a). Fig.~\ref{fig:other_td}b
shows the results for an extremely short gas dissipation time of $\tau_{\rm d} =
1$\,Myr. 
The parameter space where the mass and accretion
timescale of Mars are reproduced is not very different from that obtained when the
dissipation time is set to higher values of 5 or 10 Myr.

Recent studies suggested that terrestrial planets were constructed from
a narrow ring between the current orbits of Earth and Venus \citep[0.7
to 1\,AU;][]{morishima08,hansen,walsh}.  In this scenario, Mars was
likely scattered from around the outer edge of the truncated
disk. 
\rev{
We thus additionally investigate the possibility of Mars formation at 1\,AU by
setting the characteristic radii of the innermost and outermost annuli 
at 1 and 2.7\,AU, respectively. }
Figure \ref{fig:cont_mass_ew}b shows the final embryo mass and the
$\varepsilon^{182} {\rm W}$ value at 1\,AU in the $r_0$ and $x$ space.
The parameter space required for the production of Mars-sized embryo is
similar to the case at 1.5\,AU.  However, larger planetesimals with $r_0
\la 40$\,km could explain the mass and accretion timescale of Mars.

A very important development of this work and other recent studies 
\citep{KT10,kobayashi10} is the recognition that erosive collisions and
fragment accretion dominate embryo growth. The treatment of
fragmentation depends on $Q_{\rm D}^*$, which affects final embryo
masses \citep{kobayashi10,kobayashi11,bromley}. 
The $Q_{\rm D}^*$ value is uncertain. 
For instance, bodies made of rubble piles have $Q_{\rm D}^*$ smaller
than what we used, thus producing smaller embryos \citep{kobayashi10}. 
With $r_0 \la 10\,$km, low $Q_{\rm D}^*$ requires larger $x$ for Mars
formation, which may rise the value of plausible initial
planetesimal radii only slightly. 

\section{Discussion}

\subsection{Comparison with previous studies}

Terrestrial planet formation was investigated by $N$-body simulations
\citep{kokubo98,kokubo00,morishima}. Our statistical simulation does not
follow the orbits of bodies individually, but reproduces well the mass
and random-velocity evolution obtained from $N$-body simulations under
the assumption of perfect accretion \citep{kobayashi10}.

For terrestrial planet formation, \citet{wetherill93} performed
statistical simulations similar to ours and obtained embryos of mass $\sim
3 \times 10^{26}$\,g after about 0.1\,Myr for $x \approx 2$ and $r_0
\approx 7\,$km, which is slightly larger than embryo masses from our
simulations ($\sim 2\times 10^{26}$\,g).  Using the fragmentation
model of \citet{wetherill93}, \citet{kenyon04a} carried out statistical
simulations for $r_0 \la 1$\,km and obtained embryos of mass about
$5\times 10^{26}$\,g at 1\,Myr for $x\approx 2$. In our simulations, the
embryo masses are of about $2\times 10^{26}$\,g at 1\,Myr for $x\approx
2$ and $r_0 = 1$\,km.  
Thus, the final embryo masses predicted in our simulations are systematically lower than those obtained by previous workers for the same sets of conditions. The difference stems from a different treatment of erosive (cratering) collisions.
In the fragmentation model of 
\citet{wetherill93}, erosive collisions with specific impact energies
smaller than $Q_{\rm D}^*$ yield less ejecta than those that are expected from
impact laboratory experiments, while these are well reproduced by the
fragmentation model developed by \citet{KT10}, which is used in the
present contribution. 

\citet{weidenschilling08} carried out simulations of terrestrial planet formation where
planetary embryos were treated as individual entities, while small bodies
were followed using mass bins. These simulations showed that smaller
initial planetesimals produced smaller planetary embryos. This tendency
is consistent with ours. \citet{chambers08} used a similar model;
starting from planetary embryos and planetesimals.  The simulations in
\citet{chambers08} followed collisional fragmentation cascades using
mass bins with a mass ratio between adjacent bins of 10. These wide
mass bins artificially slowed down planetesimal depletion by the
fragmentation cascade by a factor of $\sim 4$. \citet{weidenschilling08} and \citet{chambers08}
considered the migration of embryos due to tidal interaction with the nebula
\citep[type I migration; e.g.,][]{tanaka}.  We briefly discuss the
effect of type I migration on terrestrial planet growth below.

\subsection{Subsequent Evolution of Mars}
\label{sc:subsequent}

From our simulations, 
tens of Mars size
embryos with orbital separation of 10 mutual Hill radii are expected to form in
the terrestrial planet forming region. 
For Mars to remain small during chaotic growth, 
\citet{morishima08} and
\citet{hansen} 
suggested that planetary embryos must have been concentrated inside
about 1\,AU. 
Such orbital concentration may be
attained by type I migration of embryos caused by gravitational
interaction with the surrounding nebula \citep[e.g.,][]{tanaka} if 
the torque cancellation due to the unsaturation of corotation torque
occurs at $\la 1$\,AU where heating by disk accretion determines disk
temperature \citep{paardekooper10,paardekooper11,kretke}. 
Mars may have evaded collisions with other embryos during chaotic growth by being scattered to its current orbit at 1.5\,AU. 
The planets resulting from simulations of chaotic growth have
orbital eccentricities higher than those of Venus and Earth
\citep{chambers96}, but these eccentricities were most likely damped by
gravitational interactions with a tenuous gas disk and/or a planetesimal swarm 
\citep{kominami,obrien,morishima08}. 
 This dynamical friction could have been achieved by small ($\la 10$\,m)
 leftover planetesimals that were accreted by Earth and Mars after
 completion of core formation \citep{schlichting}. The mass of the late
 veneer on Mars is estimated from abundances of highly siderophile
 elements in SNC meteorites to be $\sim 1.6\times 10^{24}$\,g. 
Assuming a chondritic impactor composition \citep[$\sim 100\,{\rm ng \, g}^{-1}$;][]{wasson},
this would have delivered $\sim 1.6 \times 10^{17}$\,g of W \citep{warren,walker}. 
The martian mantle contains $\sim 3.2 \times 10^{19}$\,g of W 
\citep[62 ppb W;][]{lodders}. 
Thus, planetesimals responsible for damping the eccentricities and
inclinations of the terrestrial planets may have delivered only $\sim 0.5\%$ of the W inventory of the martian mantle and would have had a negligible effect on the W isotopic composition.


\citet{walsh} proposed that Jupiter migrated inward until Saturn was caught in a 2:3 resonance with Jupiter, at which point Jupiter migrated outwards. The U-turn migration of Jupiter at 1.5 AU would have truncated the disk and could explain Mars present orbit and mass.
If the Jupiter's passing fully cleaned up small bodies, proto-Mars
could no longer grow after it was scattered to the current orbit. 
Our simulations can be interpreted in light of this scenario by
considering Mars' isolation time $\tau_{\rm i}$, after which the embryo
stops growing. 
The U-turn migration of Jupiter occurred at $\tau_{\rm d}$
\citep{walsh}. Proto-Mars could be scattered to its current orbit
after the migration of Jupiter; $\tau_{\rm i} \ga \tau_{\rm d}$. 
For $\tau_{\rm i} \ga 20$\,Myr -- several times longer than the gas
dissipation time $\tau_{\rm d}$, 
Mars' small mass and rapid accretion is best explained by starting with small planetesimals as was discussed above (Fig.~\ref{fig:ti}a). 
However, if one considers a rapid isolation time, which is more likely
in the context of the Grand Tack scenario \citep{walsh}, models that
start with large planetesimals are allowed
(Fig.~\ref{fig:ti}b), because Mars avoids late accretion of chondritic
material. 
As discussed in Sect.~3.3, considering the conditions needed to form Jupiter and Saturn, the scenario of formation of Mars from large planetesimals in a MMSN disk is unlikely.

A standing question is whether collisions between Mars and other embryos
during chaotic growth could have lead to further growth and modified the
W isotope evolution of the martian mantle. During such collisions, the
core of the impactor may partially merge with the target core without
complete isotopic equilibration with the target mantle \citep{dahl}.
Therefore Mars growth via collisions of sub-Mars embryos in the
subsequent stage of chaotic growth might be allowed to some extent.
However, a planetesimal disk leading to small embryos would not have
contained enough mass to produce the terrestrial planets, which makes it
unlikely that Mars grew by the protracted accretion of small embryos.

Mars has experienced at least one collision at a later stage.  The
northern hemisphere of Mars is lower and has thinner crust that the
southern hemisphere, a feature that is commonly referred to as the
martian hemispheric dichotomy. Several interpretations have been
proposed, including one that calls for a large impact that would have
obliterated the northern hemisphere less than 100 Myr after Solar System
formation \citep[e.g.,][]{wilhelms,andrews,marinova,nimmo}.
\citet{marinova} used smooth particle hydrodynamics simulations to
constrain the impactor to a 1,000\,km-radius body that would have
impacted Mars at a $\sim 45^\circ$ angle. Using the mass-radius
relationship $R/R_\oplus=(M/M_\oplus)^{0.306}$ \citep[$R_\oplus$ is
Earth radius; ][]{sotin}, such an impactor would have delivered $\approx
2 \%$ of the mass of Mars. Most likely, this would have taken place
after most $^{182}$Hf had decayed, possibly $\approx 50$\,Myr after
solar system formation. We have evaluated the impact of such a late
impact on the W isotopic evolution of the martian mantle for one of the
simulations that reproduces the mass and $\varepsilon^{182}$W values of
Mars ($x=2.7$ and $r_0=3.7$\,km). The calculated $\varepsilon^{182}$W
shifts from 2.7 to 2.5 when taking into account this late impactor. This
is well within uncertainties, so our conclusions are not affected by the
possibility that the martian dichotomy was produced by a late
impactor. The assumption of a 1000-km radius body is conservative as
\citet{nimmo} considered a head on collision with a smaller body
of 320\,km radius, which would have a negligible effect on
$\varepsilon^{182}$W.  

\subsection{Radial Gradient of planetesimal sizes}
\label{141757_4Sep12}

As discussed in \S \ref{sc:subsequent}, two scenarios can explain the
small mass and $\varepsilon^{182}$W value of Mars: (i) small
planetesimals in a massive disk formed Mars or (ii) large planetesimals
in an $\sim$MMSN disk formed Mars but its growth was arrested early by inward then outward migration of Jupiter. Further constraints bearing on this issue can be obtained by examining the conditions necessary for giant planet formation.
If an embryo reaches a critical core mass, it cannot sustain an
static atmosphere and starts rapid gas accretion to form a gas giant
planet, such as Jupiter and Saturn
\citep[e.g.,][]{mizuno,ikoma}. However, before the embryo attains the
critical core mass, collisional fragmentation induced by gravitational
stirring stalls the growth of the embryo
\citep{kobayashi10}. \citet{inaba03} suggested that fragment
accretion enhanced by tenuous atmospheres allowed formation of massive
cores. \citet{levison} meanwhile pointed out that 
combination of gas drag and stirring by an embryo 
inhibits the accretion of fragments with 
radii larger than 30\,m onto the embryo. However, the typical fragment radius is smaller
than 30\,m and hence the obstacle to fragment accretion is
negligible \citep[see the discussion in][]{kobayashi10}.  
Since the removal of fragments due to radial drift is more effective
than accretion,
embryos cannot reach the critical core mass starting from small initial
planetesimals.  Although the core formation must be completed before gas
depletion, the formation timescale of a core is longer than $\tau_{\rm
d}$ for large planetesimals, for which initial runaway growth or cores is slow. 
Therefore, Jupiter's core is likely to have formed from
accretion of intermediate size planetesimals in a massive disk \citep{kobayashi11}.

Jupiter's gas accretion was truncated by its gap opening. At the edge of
the gap, a radial pressure maximum was created. Small fragments accumulated
at the pressure maximum, which might have lead to the formation of Saturn's core
in $\sim 10^6$yr \citep{kobayashi12}. 
Heat released by solid accretion stabilizes atmospheres of cores and 
delays the rapid gas accretion to form gas giant
planets, and hence high solid accretion rates of cores result in large cores \citep[e.g.,][]{ikoma}. 
Saturn's core is estimated to be
larger than Jupiter's core \citep{saumon}, 
which is difficult to explain in classical core formation models. 
\citet{lambrechts} recently proposed 
a core formation scenario by accretion of sub-meter sized pebble, which
also produces a higher accretion rate for Jupiter 
unless the pebble surface density has a flat radial profile (the radial
slope $d \ln \Sigma_{\rm s} / d \ln a > -0.5$). A promising model to explain
Saturn's large core relative to Jupiter is if Saturn's core was formed
at the gap opened by Jupiter \citep{kobayashi12}. 

In the context of the model of \citet{kobayashi11,kobayashi12}, which uses the same prescriptions as used in the present contribution, Jupiter and Saturn can be formed in limited conditions for planetesimal
sizes and disk masses. 
Large planetesimals $r_0 \approx
30$--200\,km in a massive disk $x \ga 3$ are required to form a massive
core of $5M_\oplus$ at
5.2\,AU for Jupiter  
\citep{kobayashi11}. On the other hand, the larger core of Saturn requires very rapid formation, which is realized by 
fragment accretion at the pressure maximum caused by Jupiter's gap opening
in the solar nebula. This yields an additional constraint described empirically by the relationship  $x \ga
\sqrt{r_0/{\rm 1\,km}}$ \citep{kobayashi12}. Both conditions for formation of the cores of Jupiter and Saturn are
satisfied for $x\ga 7$ and $r_0 \approx 30$--100\,km (see
Fig.~\ref{fig:cont_mass_ew}). 

As shown in Figs.~\ref{fig:cont_mass_ew}--\ref{fig:ti}, the conditions required to explain the accretion of the cores of Jupiter and Saturn are inconsistent with those inferred for Mars. The condition required for
Jupiter and Saturn formation produces Earth-sized or larger embryos in
the terrestrial planet region and the predicted $\varepsilon^{182}
W_{\rm mantle}$ adopting $f_{\rm mantle}^{\rm Hf/W} = 12$ for Earth's
mantle 
is $\ga 10$, which is much larger than $\varepsilon^{182}
W_{\rm mantle} = 1.85$ in terrestrial rocks \citep{jacobsen,kleine,dauphas2}. 

One way to try to reconcile the apparently inconsistent conditions inferred for Mars on the one hand and Jupiter and Saturn on the other hand is to modify the disk surface density. In the discussion above and in Fig.~\ref{fig:cont_mass_ew}--\ref{fig:ti}, we assumed $\Sigma_{\rm g} \propto \Sigma_{\rm s} \propto a^{-1.5}$. It is possible for instance that the disk surface density had a shallower radial slope of $\Sigma_{\rm g} \propto \Sigma_{\rm s} \propto a^{-1}$.  The conditions for
forming the cores of Jupiter and Saturn become $x\ga 2.4$ and $r_0 \ga
40$\,km. If we use a massive disk of 0.1 solar mass required by 
Jupiter formation, 
the radius of initial planetesimals for Mars
formation becomes less than 5\,km for $\Sigma_{\rm g} \propto \Sigma_{\rm s}
\propto a^{-1}$. 
Changing the disk density profile does not seem to solve the discrepancy
that exists between Mars and the giant planets.

A way to reconcile the constraints derived from inner and outer solar
system bodies is if planets formed in a relatively massive solar nebula
($x \ga 5$) but planetesimals sizes were different in the inner and
outer regions. In this scenario, Jupiter and Saturn cores would have
formed from planetesimals of 50\,km radius while Mars formed from
smaller planetesimals of $\la 10$ km radius. \citet{morbidelli} proposed
that 100\,km or larger planetesimals were necessary to explain the mass
distribution of asteroids in the main belt, but
\citet{weidenschilling11} concluded that planetesimals of 100\,m size
could also reproduce the asteroid mass distribution.  The sizes of
planetesimals are related to their formation mechanisms \citep[{\it
e.g.},][]{okuzumi,johansen}.  
The heliocentric gradient in planetesimal sizes discussed here may be related to the inferred contrast in solid surface densities and material properties across the snow line.

\vspace{1cm}

We thank D. Minton, A. Morbidelli, and an anonymous reviewer for 
comments that helped improve the manuscript. 
HK gratefully acknowledges the support from 
Grants-in-Aid from MEXT (23103005) and "Institutional Program for Young
Researcher Overseas Visits" (R29).
ND acknowledges the support of NASA (NNX12AH60G) and NSF
(EAR-0820807). 

\appendix

\section{Derivation of $\varepsilon ^{182}$W evolution}
\label{appendix}

The mantle $^{182}$W value evolves due to extraterrestrial
delivery, removal of metal to core, and radioactive decay of $^{182}$Hf. 
A planetary embryo mainly grows through collisions with small bodies. If
the sizes of bodies are much smaller than the thickness of the embryo
mantle, the sinking metal breaks up in tiny droplets that can
equilibrate isotropically with the surrounding mantle \citep{rubie,samuel}. 
On the other hand, 
one third of embryo mass growth is provided by collisions with similar
mass bodies \citep{chambers06,kobayashi10,kobayashi11}. 
According to \citet{morishima}, 
contribution of large bodies to the $\varepsilon^{182}$W
evolution of the martian mantle is negligible. 
Therefore, we assume that the bodies accreted 
by Mars are fully equilibrated with the mantle before metal removal into the core.
Accordingly, the equation governing the $^{182}$W isotopic evolution is,
\begin{equation}
 \frac{d}{dt}(M_{\rm mantle} \, [^{182}{\rm W}]_{\rm mantle}) = 
\, [^{182}{\rm W}]_{\rm CHUR} \frac{dM}{dt} 
-D^{\rm W} \, [^{182}{\rm W}]_{\rm mantle} \frac{d M_{\rm core}}{dt}
+ \lambda \, [^{182}{\rm Hf}]_{\rm mantle} M_{\rm mantle},\label{142100_5Nov12} 
\end{equation}
where $M_{\rm mantle}$ and $M_{\rm core}$ are, respectively, the mantle
and core masses ($M = M_{\rm mantle}+M_{\rm core}$), 
$D^{\rm W}$ is the W metal/silicate partition coefficient. 
The mass balance for W in the bulk planet can be written as,
\begin{equation}
 [{\rm W}]_{\rm mantle} M_{\rm mantle} + D^{\rm W} [{\rm W}]_{\rm mantle} M_{\rm core} = [{\rm W}]_{\rm CHUR} M.\label{124541_5Nov12} 
\end{equation}
On the other hand, Hf does not go to the mantle and the mass balance is
given by 
\begin{equation}
 [{\rm Hf}]_{\rm mantle} M_{\rm mantle} = [{\rm Hf}]_{\rm
  CHUR} M.\label{125138_5Nov12} 
\end{equation}
We assume that the core mass fraction, $\gamma = M_{\rm core}/M$, is constant. 
We then obtain, from Eqs.~(\ref{124541_5Nov12}) and
(\ref{125138_5Nov12}), 
\begin{equation}
 f_{\rm mantle}^{\rm Hf/W} 
\equiv \frac{({\rm Hf}/{\rm W})_{\rm mantle}}{({\rm Hf}/{\rm W})_{\rm CHUR}} -1 
= \frac{D^{\rm W} \gamma}{1-\gamma}.\label{144436_5Nov12} 
\end{equation}

The variation of $R_{\rm m}= (^{182}W/^{184}W)_{\rm mantle}$ and $R_{\rm
CHUR} = (^{182}W/^{184}W)_{\rm CHUR}$ is given by 
Eq.~(\ref{142100_5Nov12}) divided by $^{184}{\rm W}_{\rm mantle} M$ using Eqs.~(\ref{124541_5Nov12}): 
\begin{eqnarray}
(1- \gamma) \left( R_{\rm m} \frac{d \ln M}{dt} + \frac{d R_{\rm m}}{dt}
	    \right) 
&=& (1-\gamma + D^{\rm W} \gamma) R_{\rm CHUR} \frac{d \ln M}{dt} 
\nonumber
\\
&&
- D^{\rm W} \gamma R_{\rm m} \frac{d \ln M}{dt} + \lambda (1-\gamma)
\left(\frac{^{182} {\rm Hf}}{^{184} {\rm W}}\right)_{\rm mantle}. 
\end{eqnarray}
Using Eq.~(\ref{144436_5Nov12}) and 
$(^{182}{\rm Hf}/^{182}{\rm W})_{\rm
mantle} =(1+f_{\rm mantle}^{\rm Hf/W}) (^{182}{\rm Hf}/^{182}{\rm W})_{\rm CHUR}$
and $dR_{\rm CHUR}/dt = \lambda (^{182}{\rm Hf}/^{184}{\rm W})_{\rm
CHUR}$, 
we then have 
\begin{equation}
 \frac{d}{dt} (R_{\rm m}-R_{\rm CHUR}) = 
  - (f_{\rm mantle}^{\rm Hf/W}+1) (R_{\rm m} - R_{\rm CHUR}) 
  \frac{d \ln M}{dt} +
  \lambda f_{\rm mantle}^{\rm Hf/W}
  \left(\frac{^{182}{\rm Hf}}{^{184}{\rm W}}\right)_{\rm CHUR}.\label{133959_23Nov12} 
\end{equation}
Integration of Eq.~(\ref{133959_23Nov12}) results in 
\begin{equation}
 R_{\rm m} - R_{\rm CHUR} = \lambda f_{\rm mantle}^{\rm Hf/W}
  \left(\frac{^{182}{\rm Hf}}{^{184}{\rm W}}\right)_{\rm CHUR,0} 
  \int_{\rm 0}^{t} \left(
		    \frac{M(t^{\prime})}{M(t)} 
\right)^{1+f_{\rm mantle}^{\rm Hf/W}} dt^{\prime}.\label{153641_5Nov12} 
\end{equation}
Applying the definition of
$\varepsilon^{182}W$, 
Eq.~(\ref{153641_5Nov12}) 
can be rewritten as 
Eq.~(\ref{eq:eW}). 




\section*{Figure Caption}

\begin{figure}[htbp]
\epsscale{1} \plottwo{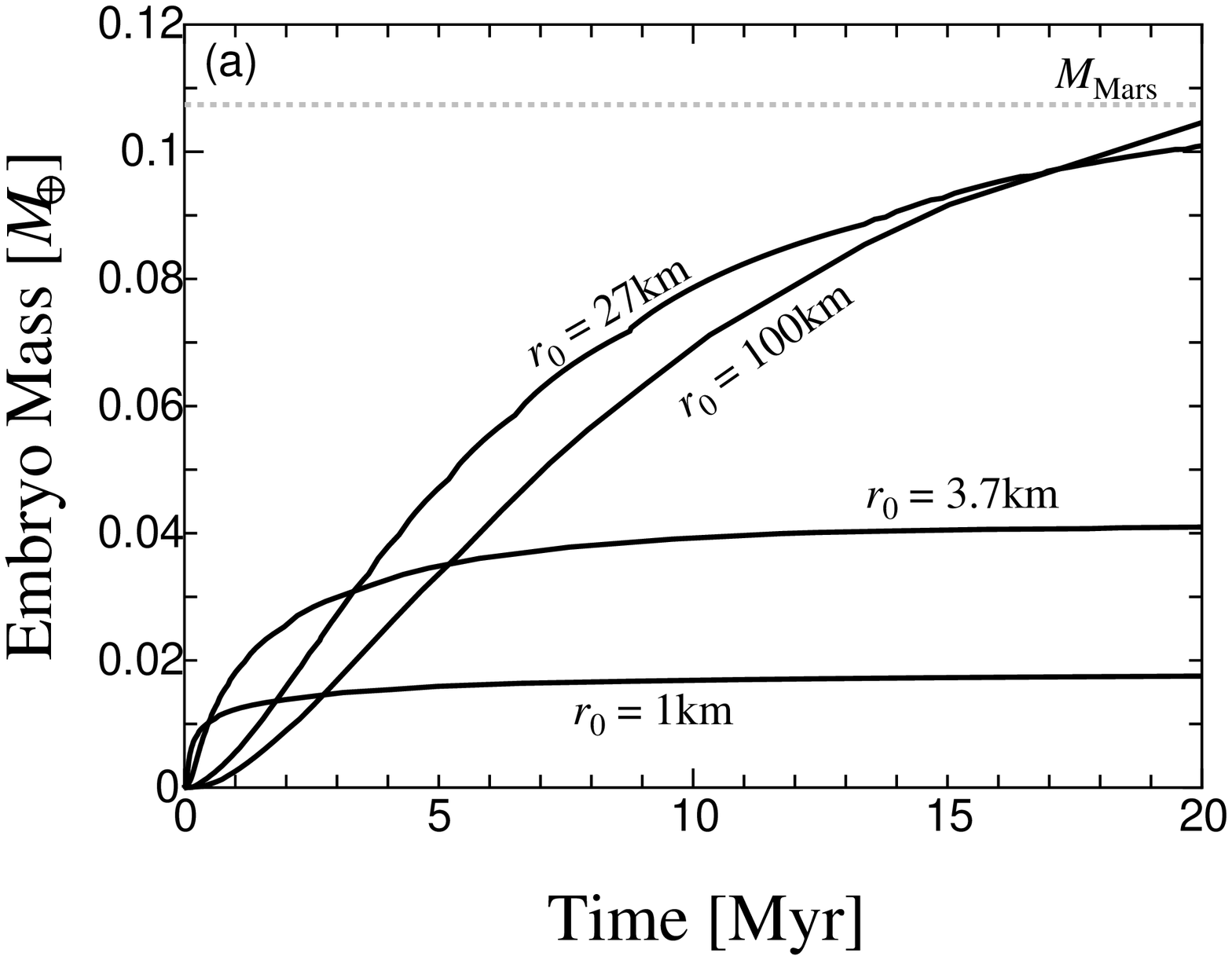}{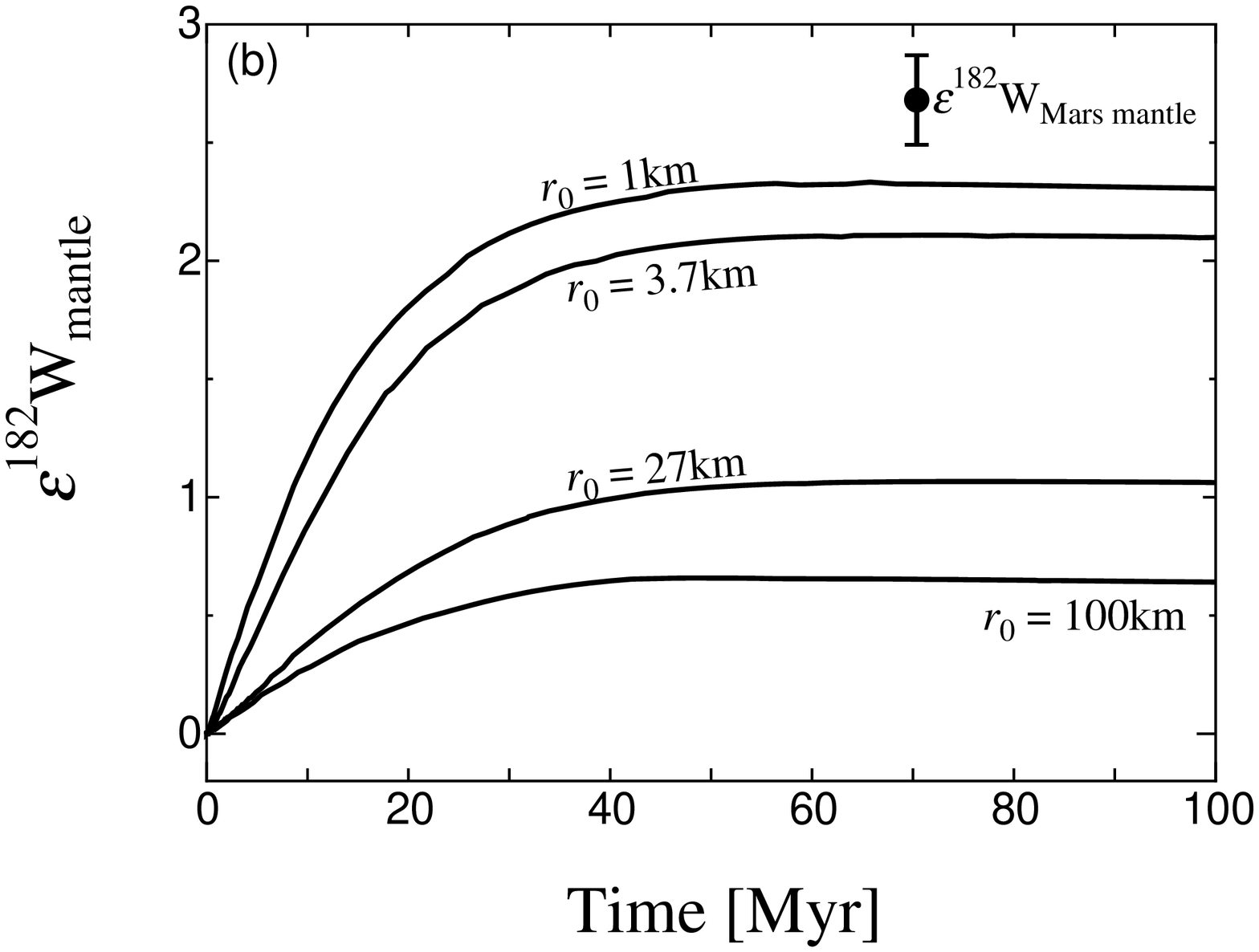} \figcaption{ 
\label{fig:embryo_growth}}
\end{figure}

Fig.~\ref{fig:embryo_growth} --- 
(a) Time
evolution of embryo mass at 1.5\,AU in the MMSN disk ($x=1$), starting from
different planetesimal radii $r_0$.  
The gray dotted line indicates the mass of Mars. 
(b)
Evolution of $\varepsilon^{182}{\rm W}_{\rm mantle}$ obtained from the mass evolution
shown in panel (a). The filled circle and the error
bar represent the $^{182}$W excess in the martian mantle, $\varepsilon^{182}{\rm
 W}_{\rm Mars\,mantle}$ \citep{kleine,dauphas}.

\begin{figure}[htbp]
\epsscale{1} \plottwo{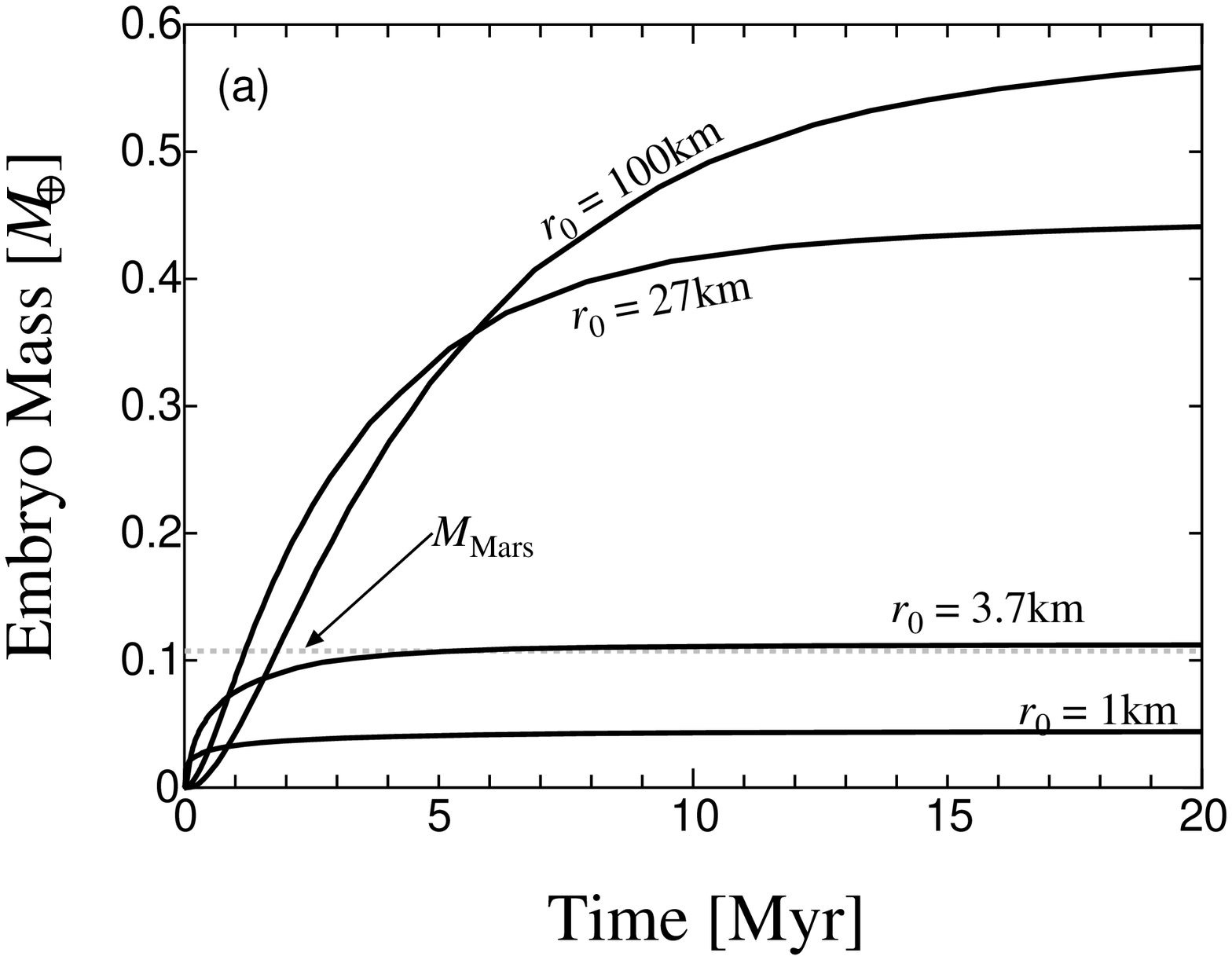}{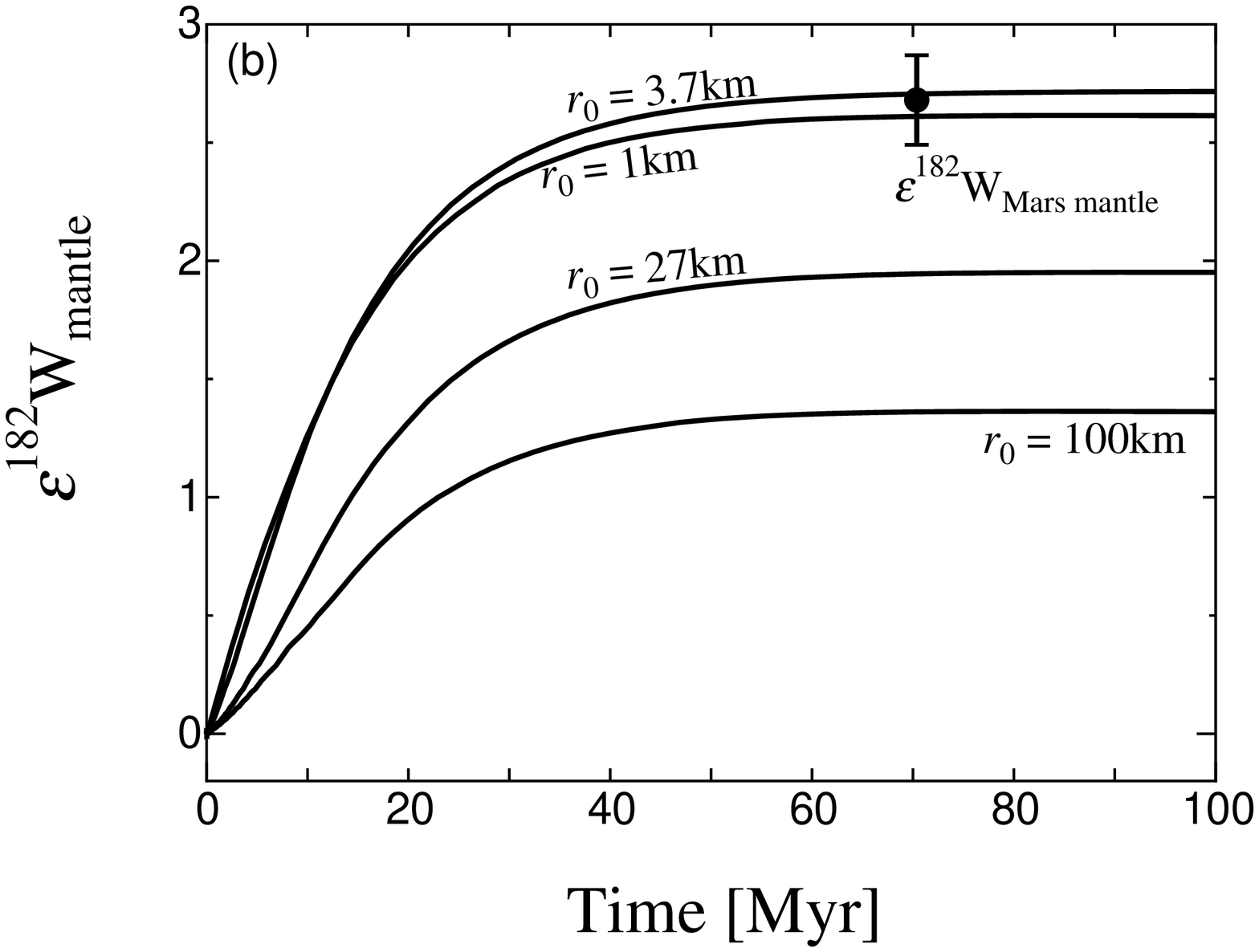} \figcaption{ 
\label{fig:embryo_growth_s2.7}}
\end{figure}

Fig.~\ref{fig:embryo_growth_s2.7} --- 
Same as Fig.~\ref{fig:embryo_growth} but for $x = 2.7$.

\begin{figure}[htbp]
\epsscale{1} \plottwo{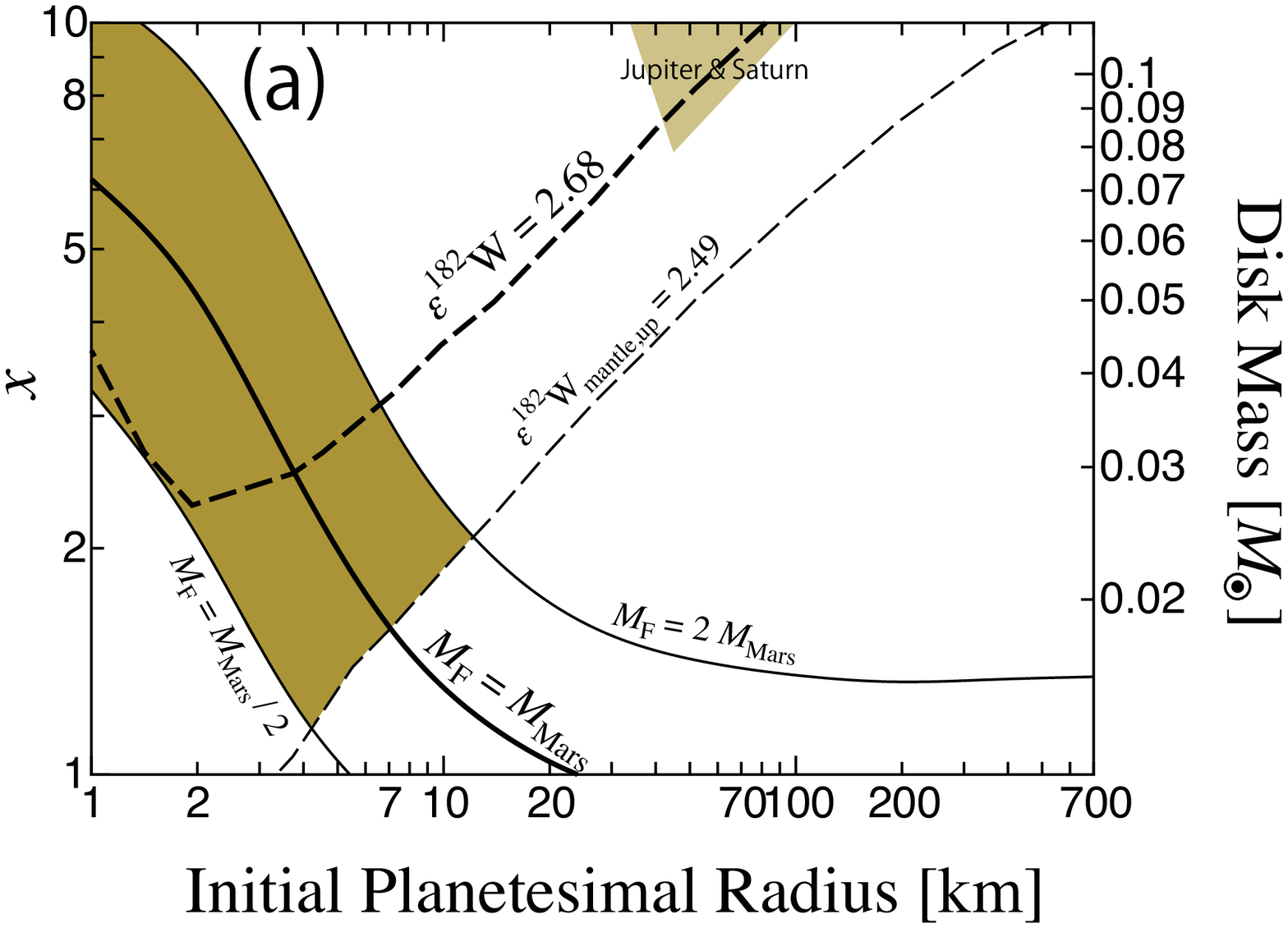}{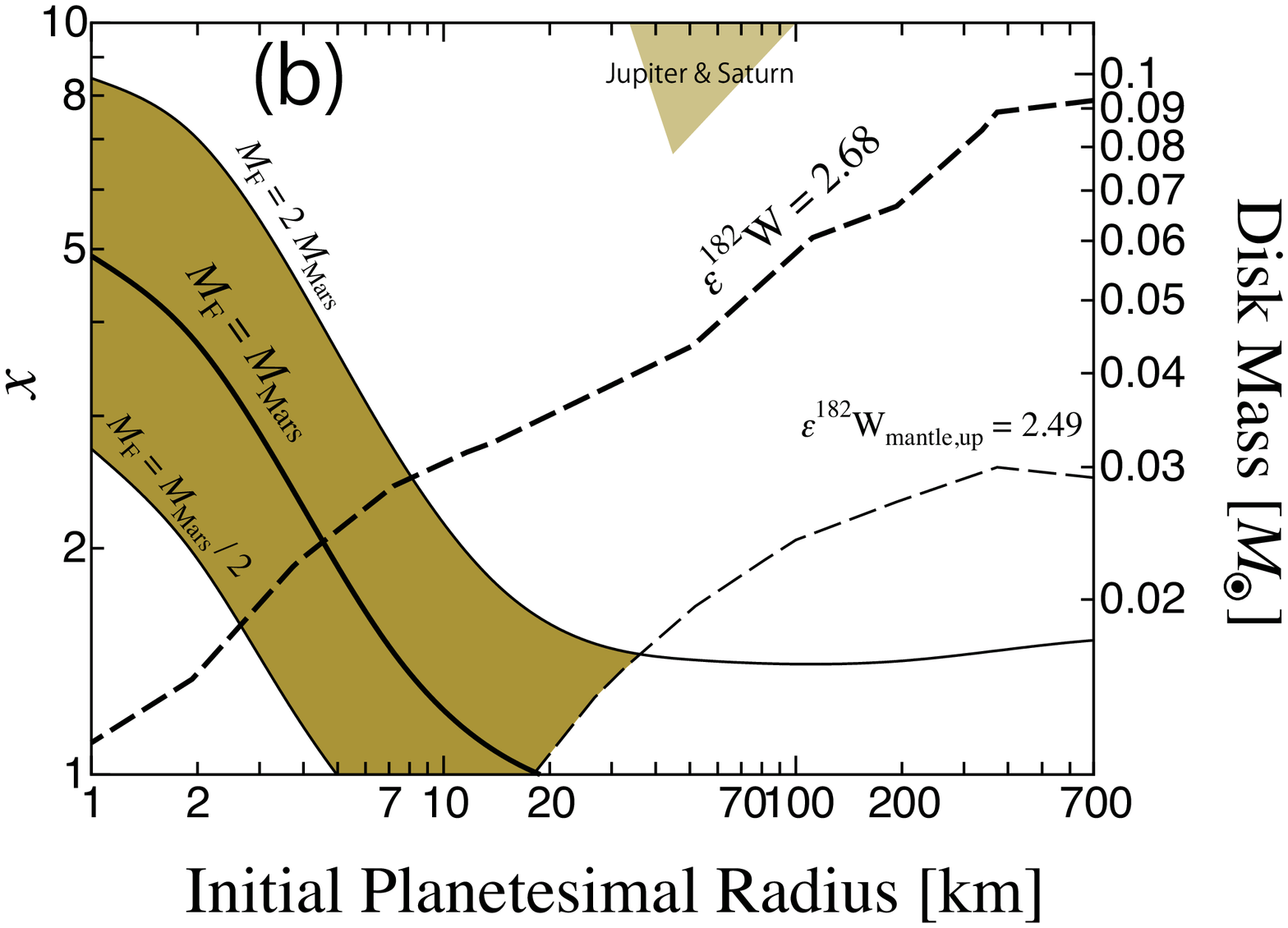} \figcaption{ 
\label{fig:cont_mass_ew}}
\end{figure}

Fig.~\ref{fig:cont_mass_ew} --- 
Constraints from final embryo masses and $\varepsilon^{182}{\rm W}$ on
the initial radius of planetesimals, $r_0$, at
 1.5\,AU (a) and 1\,AU (b) in the $x\times$MMSN disk. 
Embryos reach the mass of Mars on the solid curve. 
The region between the thin curves produces embryos within a factor 2 of
 the mass of Mars. 
 On the thick dashed curve, the predicted $\varepsilon^{182}{\rm W}_{\rm
 mantle}$ values
 correspond to the mean value obtained from Martian meteorites. 
The thick dashed curve indicates the parameter space where 
 $\varepsilon^{182}{\rm W}_{\rm mantle,up}
 = \varepsilon^{182}{\rm W}_{\rm
 mantle} + \Delta \varepsilon^{182}{\rm W}_{\rm mantle} 
= 2.49$. Below this limit, the accretion timescale of Mars would be too long to explain  $\varepsilon^{182}{\rm W}_{\rm mantle}$ measured in Martian meteorites.
The colored region at small $r_0$ represents the condition satisfying 
the mass and $\varepsilon^{182}{\rm W}_{\rm mantle}$ value of Mars and 
the other at $r_0 \approx 40$--100\,km indicates the parameters for the
formation of Jupiter and Saturn cores. 
\revs{
For reference, the right vertical axis represents the disk mass
corresponding to $x$ in the left axis, assuming that the inner and outer
edges of the disk are at 0.4 and 30\,AU, respectively. 
}

\begin{figure}[htbp]
\epsscale{1} \plottwo{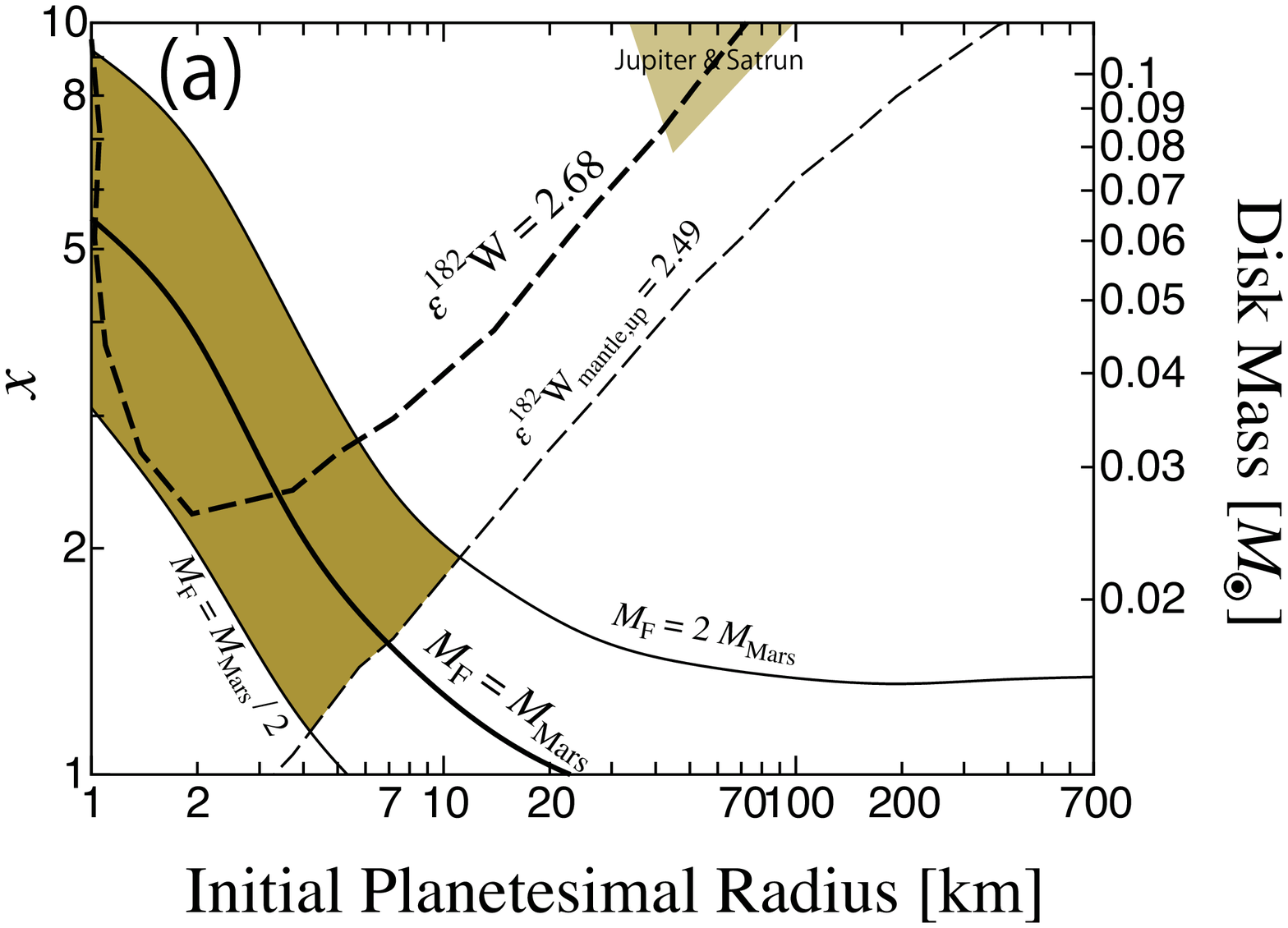}{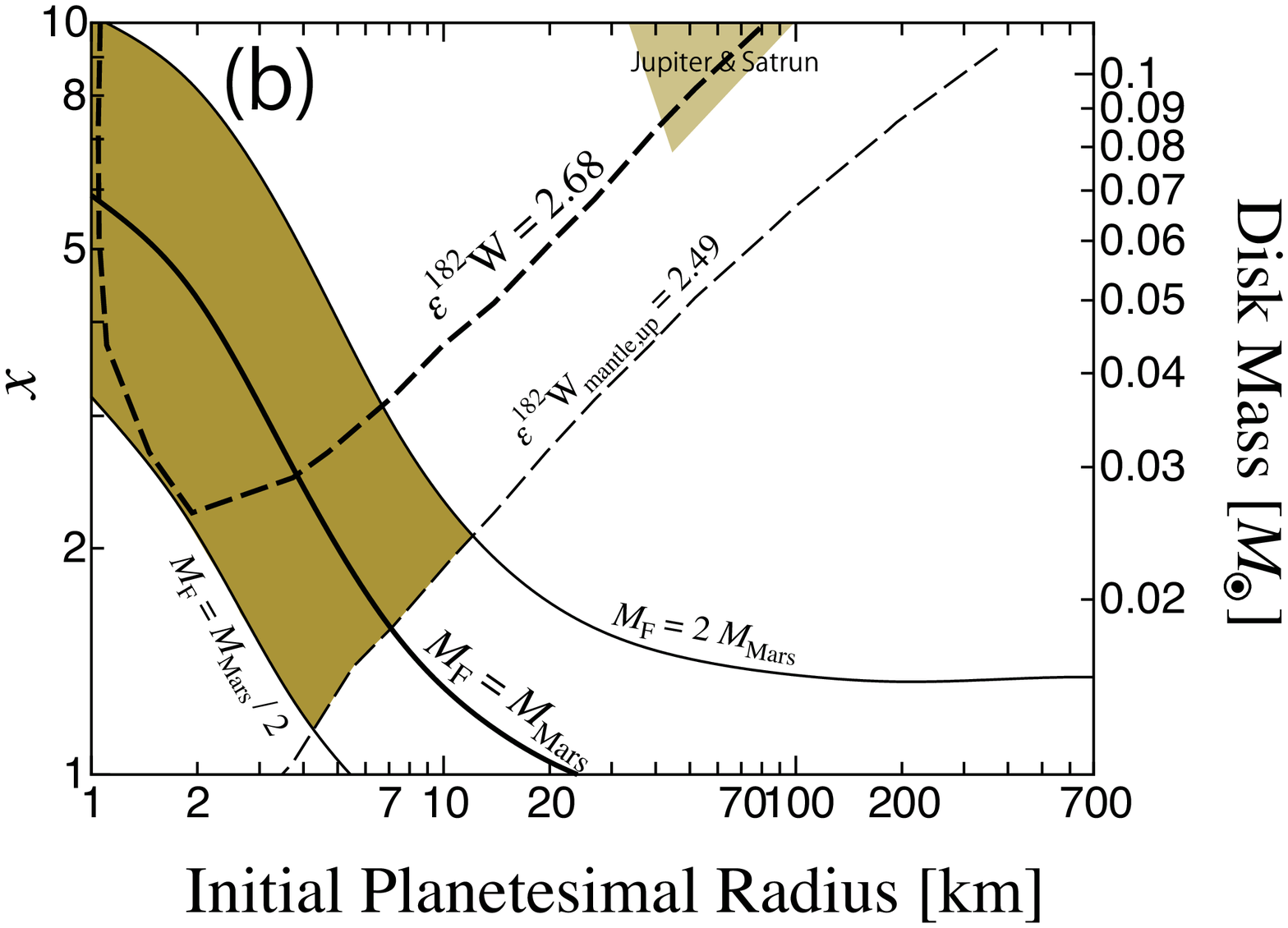} \figcaption{ 
\label{fig:other_effects}}
\end{figure}

Fig.~\ref{fig:other_effects} ---
Same as Fig.~\ref{fig:cont_mass_ew}a, but including the accretion
enhancement by planetary atmosphere (a) and gas drag (b). 

\begin{figure}[htbp]
\epsscale{1} \plottwo{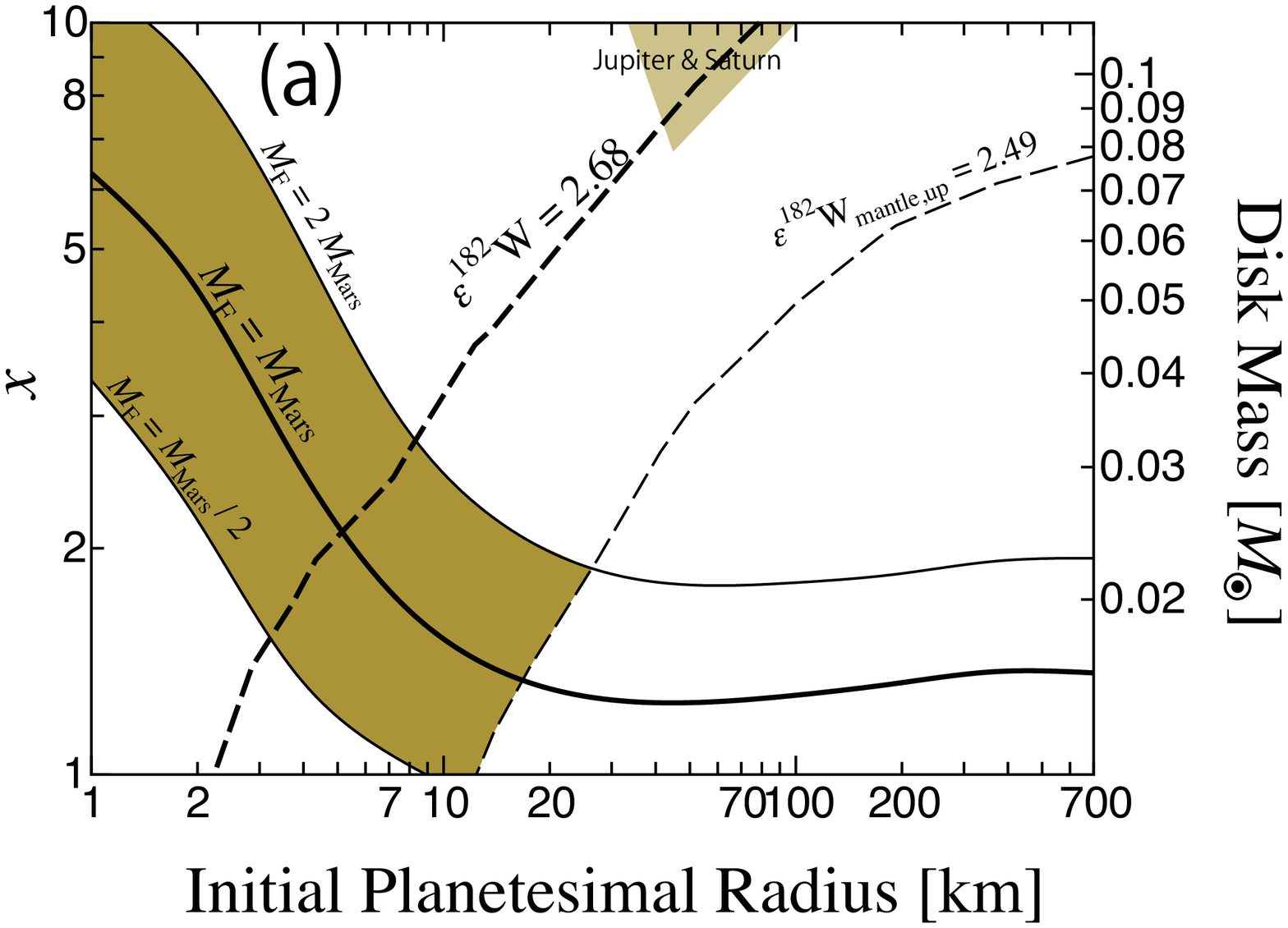}{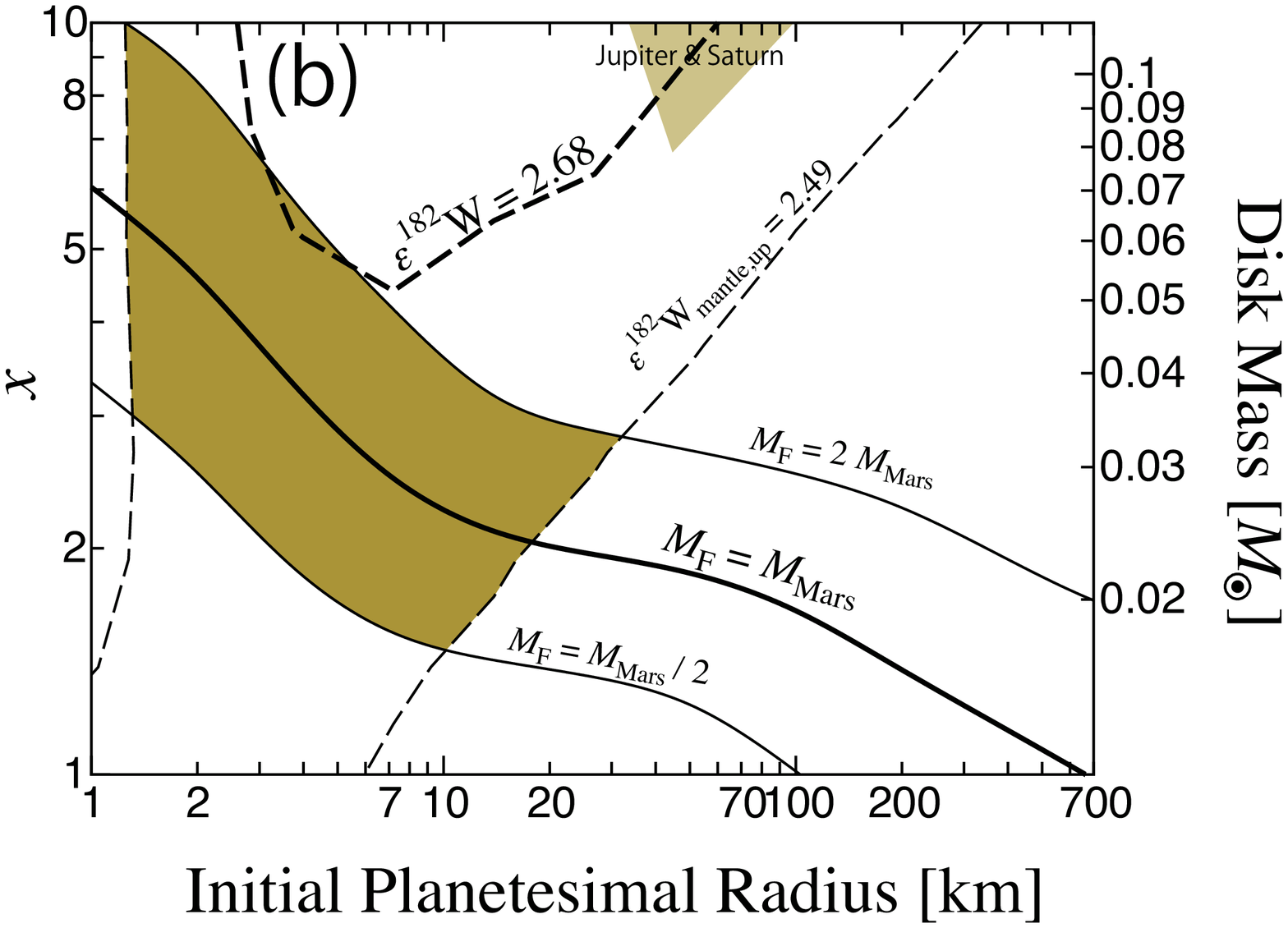} \figcaption{ \label{fig:other_td}}
\end{figure}

Fig.~\ref{fig:other_td} ---
\rev{
Same as Fig.~\ref{fig:cont_mass_ew}a, but for different gas dissipation
timescales $\tau_{\rm d} = 5$\,Myr
(a) and 1\,Myr(b). 
}

\begin{figure}[htbp]
\epsscale{1} \plottwo{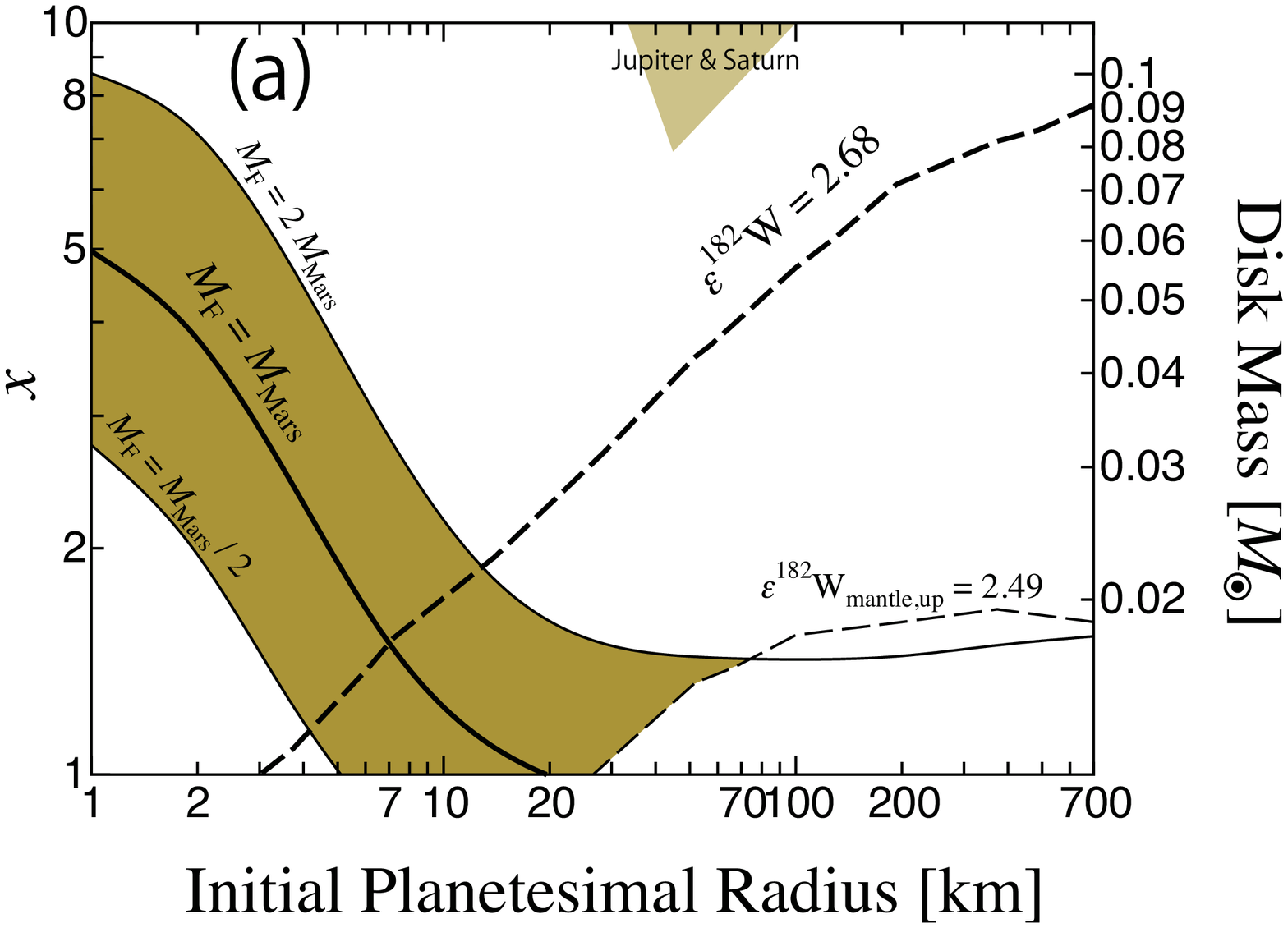}{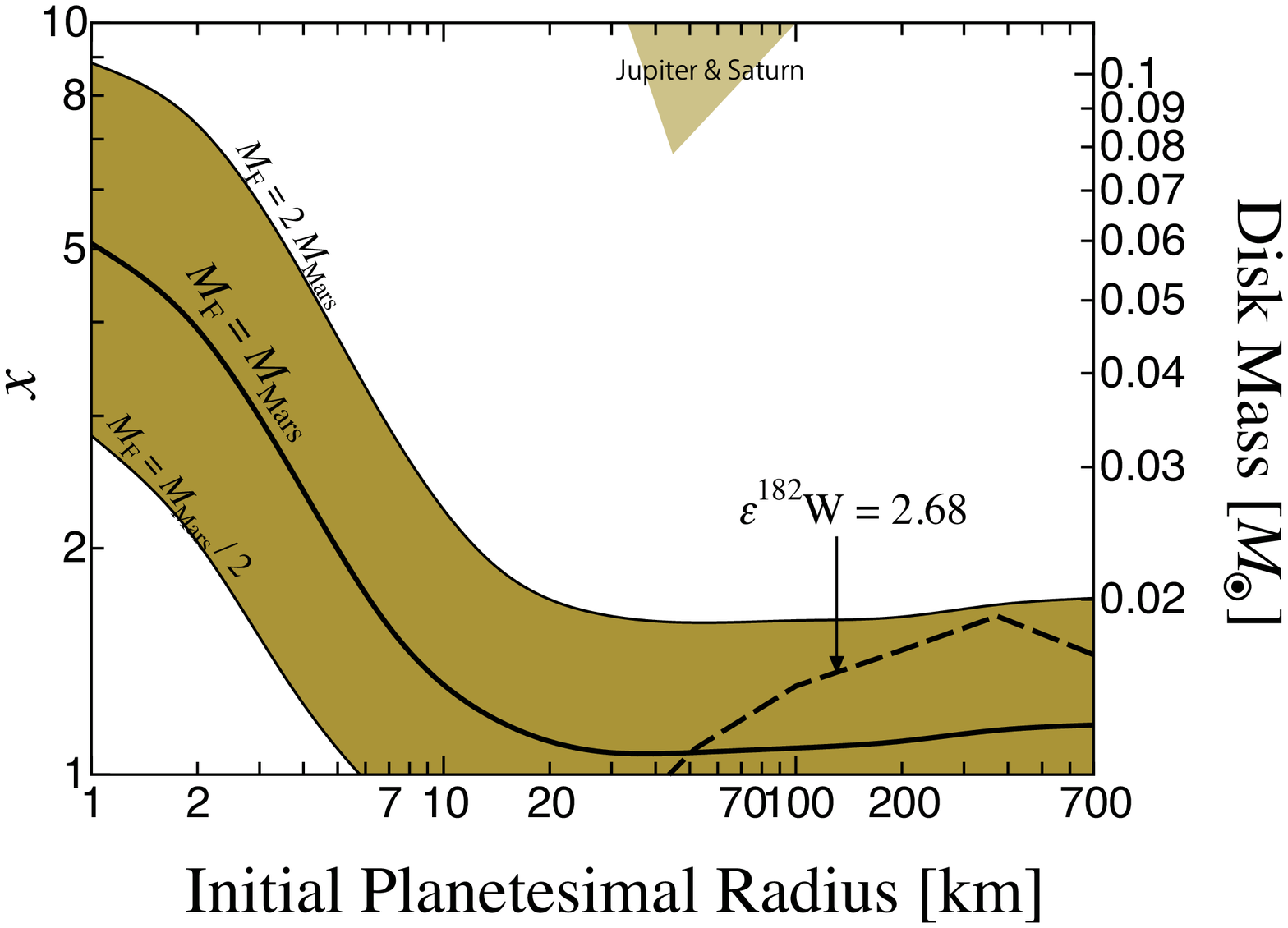}
 \figcaption{\label{fig:ti}}
 \end{figure}

Fig.~\ref{fig:ti} --- 
Same as Fig.~\ref{fig:cont_mass_ew}b (at 1\,AU), but introducing the
isolation time of mars, $\tau_{\rm i} = 20$\,Myr (a), 5\,Myr (b), after
which Mars does not grow at all. 

\newpage 

\end{document}